\documentclass[epj]{webofc}
\usepackage[varg]{txfonts}   
\usepackage{hyperref}
\usepackage{url}
\usepackage{lineno}
\hypersetup{colorlinks=true,citecolor=blue,urlcolor=blue,linkcolor=blue}
\begin{document}
\title{Accelerating ALICE Background Simulations: A WGAN-GP Approach to
Slow-Neutron-Induced TPC Loopers}
\author{
  \firstname{Marco} \lastname{Giacalone}\inst{1}\fnsep\thanks{\email{marco.giacalone@cern.ch}}
  \and
  \firstname{Sandro Christian} \lastname{Wenzel}\inst{1}
  \and
  \firstname{Mesut} \lastname{Arslandok}\inst{2}
  \and
  \firstname{Andreas} \lastname{Morsch}\inst{1}
}

\institute{
  CERN, Geneva, Switzerland
  \and
  Yale University, New Haven, CT, USA
}

\abstract{
The ALICE experiment is targeting to produce the most realistic
description of particle production by estimating accurately the reconstruction efficiency in its Monte Carlo simulations. This
requires including background contributions that are often disabled by
default because of their computational cost. One of these contributions
originates indirectly from slow neutrons interacting with the enclosure of the Time
Projection Chamber (TPC): the resulting low-momentum electron-positron
pairs and Compton electrons spiral repeatedly in the detector
magnetic field, producing the ``loopers'' that account for
approximately 10\% of all reconstructed TPC clusters. Transporting the parent slow neutrons through GEANT4
down to capture is very expensive, so a generative machine-learning
approach based on a Wasserstein Generative Adversarial Network with
gradient penalty (WGAN-GP) has been
developed to speed up the process. It was trained on labelled samples from full
Pythia8+GEANT4 simulations, and directly generates the final-state
pair-production and Compton-scattering particles, bypassing the
slow-neutron transport entirely. The model is integrated into the
ALICE O2 simulation framework via ONNX Runtime and is now used in
production for pp and Pb--Pb simulations. It provides results consistent
with full simulation, reproducing the looper occupancy observed in the TPC, and it achieves
a measured speed-up of about a factor ten in the particle transport step.
}
\maketitle
\section{Introduction}
\label{sec-introduction}
A Large Ion Collider Experiment (ALICE)~\cite{Alice2008,ALICE:2023udb} is one of the
main experiments at the LHC at CERN, dedicated to the study of the
quark--gluon plasma in heavy-ion collisions. Extracting physics
results at the required precision relies on detailed Monte Carlo (MC)
simulations that reproduce the experimental data as closely as
possible. Achieving this level of
realism requires the description of a large number of physics contributions
that are generally disabled by default in event generators and
transport frameworks (GEANT4) because of their computational cost, making
simulations time-consuming. These contributions let us model better the detector occupancy, leading to a more precise estimate of the reconstruction efficiency and resolution. To mitigate this
cost, ALICE is making use of machine learning (ML) techniques to
accelerate selected simulation steps without sacrificing physics
accuracy.

In this paper, the fast ML-based simulation of TPC ``loopers'' produced by slow-neutron interactions
with the TPC enclosure is described. The clusters generated by these particles in the detector were studied with dedicated Monte Carlo simulations 
and they were found to be approximately 10\% of all reconstructed ones during data taking, and are therefore an important
ingredient of any realistic background description. We introduce the detector in which loopers are observed and the physics behind them
(Sects.~\ref{sec-tpc}-\ref{sec-loopers}), explain how a labelled training sample is
obtained from full simulation (Sect.~\ref{sec-tagging}), 
detail the generative model (Sect.~\ref{sec-ml}), describe its
integration into the ALICE O2 software framework
(Sect.~\ref{sec-integration}), and present the resulting performance
and physics validation (Sect.~\ref{sec-results}).

\section{The ALICE Time Projection Chamber}
\label{sec-tpc}
\begin{figure}[h]
\centering
\includegraphics[width=7.5cm,clip]{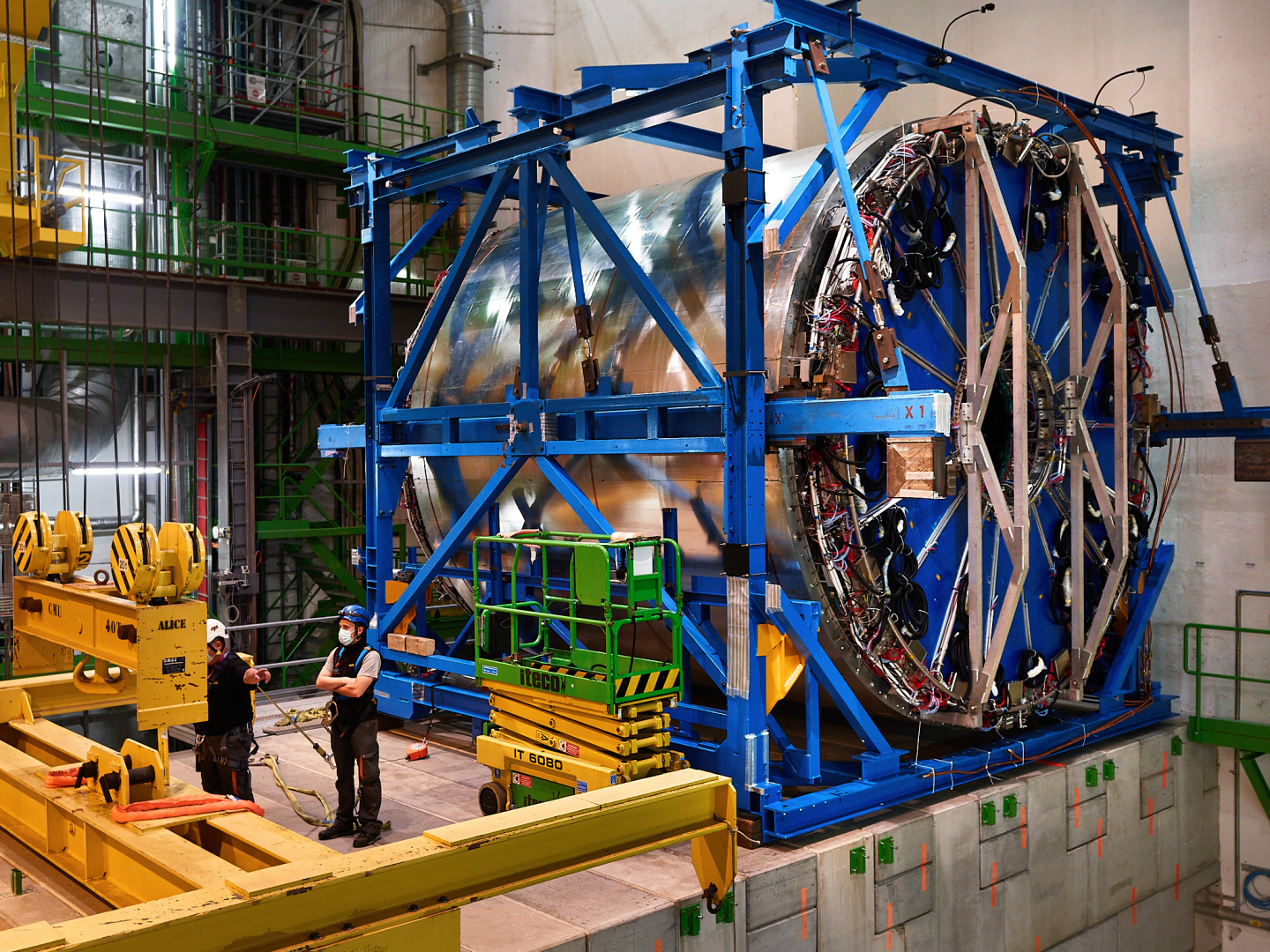}
\caption{The ALICE TPC during its re-installation [CERN-PHOTO-202008-104-85] in the experimental
cavern after the LS2 upgrade; people are shown for scale.}
\label{fig-tpc-installation}
\end{figure}

The Time Projection Chamber is one of the main tracking and
particle-identification detectors of the ALICE experiment. With a gas
volume of approximately 88~m$^3$, it is the largest detector of its kind ever
built~\cite{Alme2010tpc,ALICETPC:2020ann}, and it is shown in
Fig.~\ref{fig-tpc-installation}, a photograph taken during its re-installation in
the ALICE cavern after its upgrade, in Long Shutdown 2 (LS2).

\begin{figure}[h]
\centering
\includegraphics[width=0.7\textwidth]{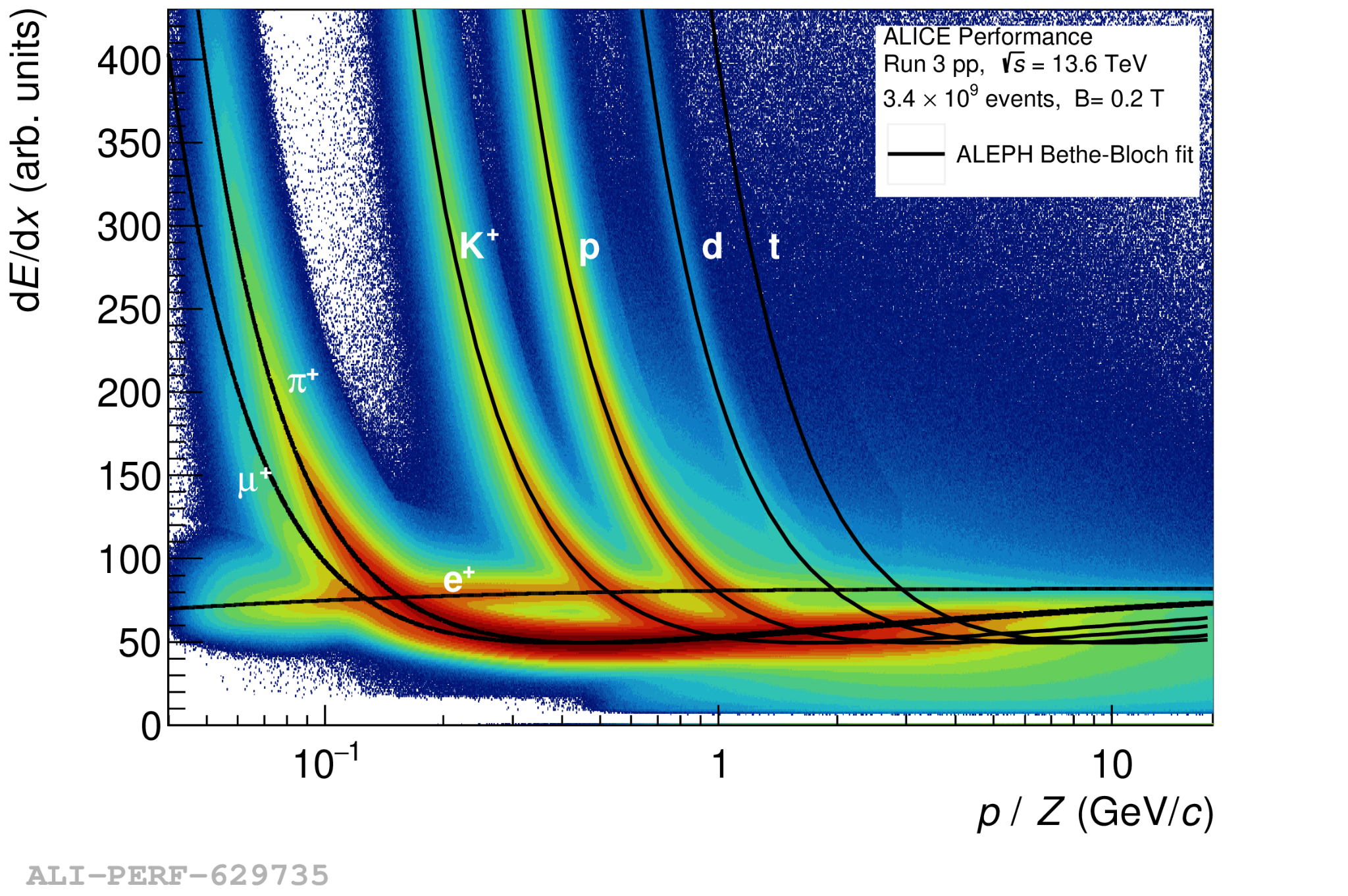}
\caption{TPC specific energy loss d$E$/d$x$ as a function of
momentum over charge, measured in Run~3 pp collisions at
$\sqrt{s}=13.6$~TeV, with an ALEPH Bethe--Bloch fit overlaid for each
particle species.}
\label{fig-tpc-dedx}
\end{figure}

The detector provides full 3D track reconstruction from the
drift of ionisation electrons in its electric field, and particle
identification (PID) via the specific energy loss d$E$/d$x$ of
charged particles traversing its gas volume. Figure~\ref{fig-tpc-dedx}
shows the d$E$/d$x$ performance obtained in Run~3 pp collisions at
$\sqrt{s}=13.6$~TeV. In the plot, the different particle species are identified by their
separated bands that are well described by an ALEPH Bethe--Bloch parametrisation \cite{ALICE:2014sbx} corrected by a neural network \cite{Sonnabend:2025uio}.

\section{TPC loopers from slow-neutron interactions}
\label{sec-loopers}
Slow neutrons produced in the collision and in secondary interactions
with the surrounding material can be captured by nuclei in the TPC
field cage and enclosure. The resulting excited nucleus de-excites by
emitting one or more $\gamma$ rays, which in turn convert in the TPC
gas predominantly through Compton scattering, giving a single low-momentum
electron, or through pair production, generating an $e^+e^-$ pair. Because of their very low momentum, these leptons repeatedly
spiral (``loopers'') within the TPC gas volume under the $0.5$~T
solenoidal magnetic field before eventually stopping or exiting the
active volume, rather than crossing it once like a typical
high-momentum track.  Two effects make them span many time bins. The
helix advances along the beam axis while its projection in the
transverse plane stays almost fixed, so a single looper deposits
charge over a wide range of drift times at nearly constant radius; and
the parent neutron must thermalise before being captured, so the
leptons are created microseconds after the collision that produced it,
outside the drift window of any single interaction.

\begin{figure}[h]
\centering
\includegraphics[width=0.99\textwidth]{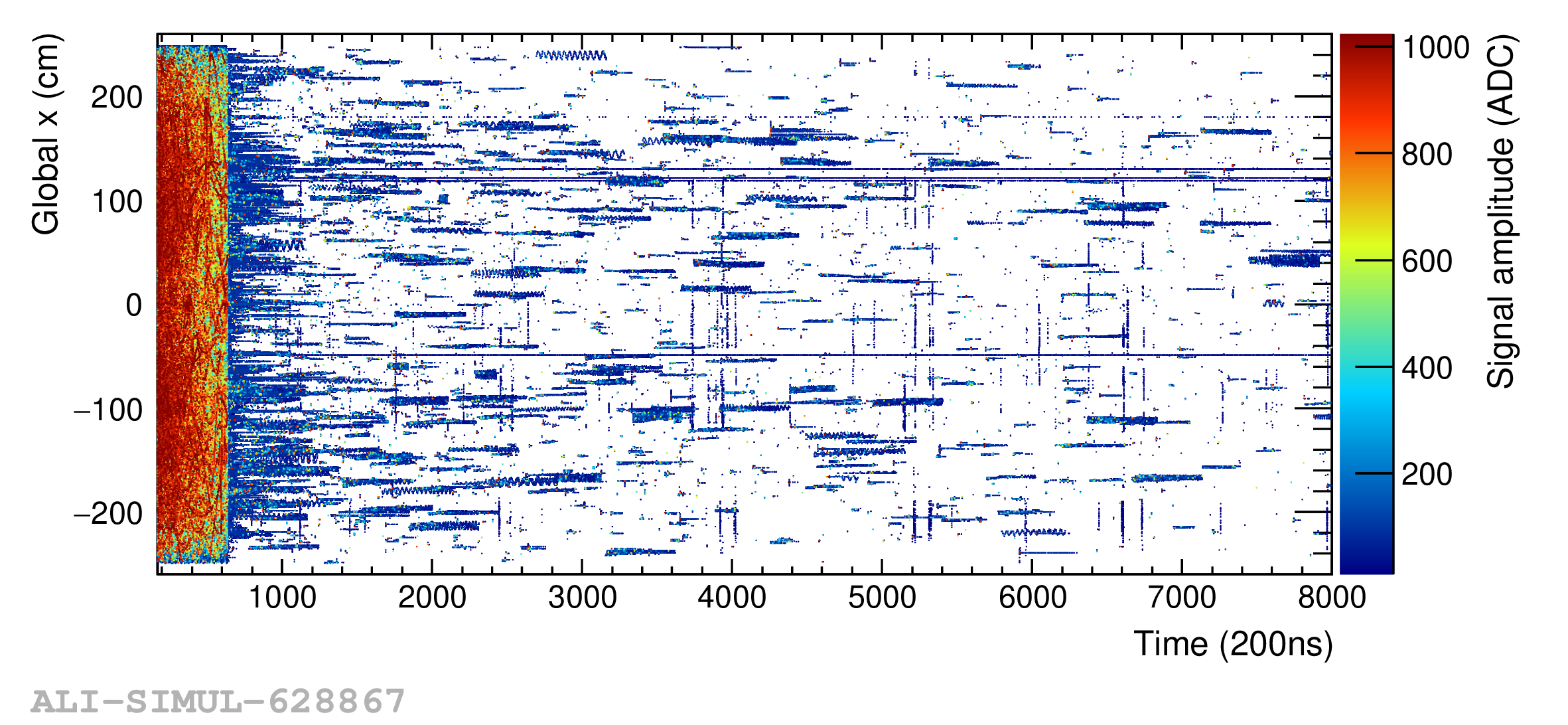}
\caption{Reconstructed TPC clusters (global $x$ position versus time in TPC units - 200 ns/bin)
in experimental Pb--Pb run. Loopers are identified by their long life after the bulk of the event.}
\label{fig-looper-sector}
\end{figure}

In reconstructed data these loopers are identified by their long-lived,
helical structures spanning many TPC readout time bins,
in clear contrast with the short-lived clusters left by primary
tracks. Figure~\ref{fig-looper-sector} shows an example of reconstructed clusters from Pb--Pb collisions in Run~3 at $\sqrt{s_{\rm NN}}=5.36$~TeV, where the many long-lived clusters
after the bulk of the event (approximately one full TPC drift for a single event, which is $\approx$90 $\mu$s) are a signature of loopers production.

\section{Training sample generation and tagging}
\label{sec-tagging}
A labelled sample of loopers was obtained from a full simulation
chain using Pythia 8~\cite{Bierlich:2022pfr} with the Angantyr~\cite{Angantyr} model as the event generator
for Pb--Pb collisions at $\sqrt{s_{\rm NN}}=5.36$~TeV, with particle
transport performed by GEANT4~\cite{Agostinelli2003geant4,Allison:2016lfl}. By
default, the transport of very low kinetic energy neutrons is
suppressed by production cuts for computational efficiency. For the purpose of building the training sample,
this cut was relaxed in a dedicated simulation campaign
so that slow-neutron transport was fully simulated.
\begin{figure}[h]
\centering
\includegraphics[width=0.5\textwidth]{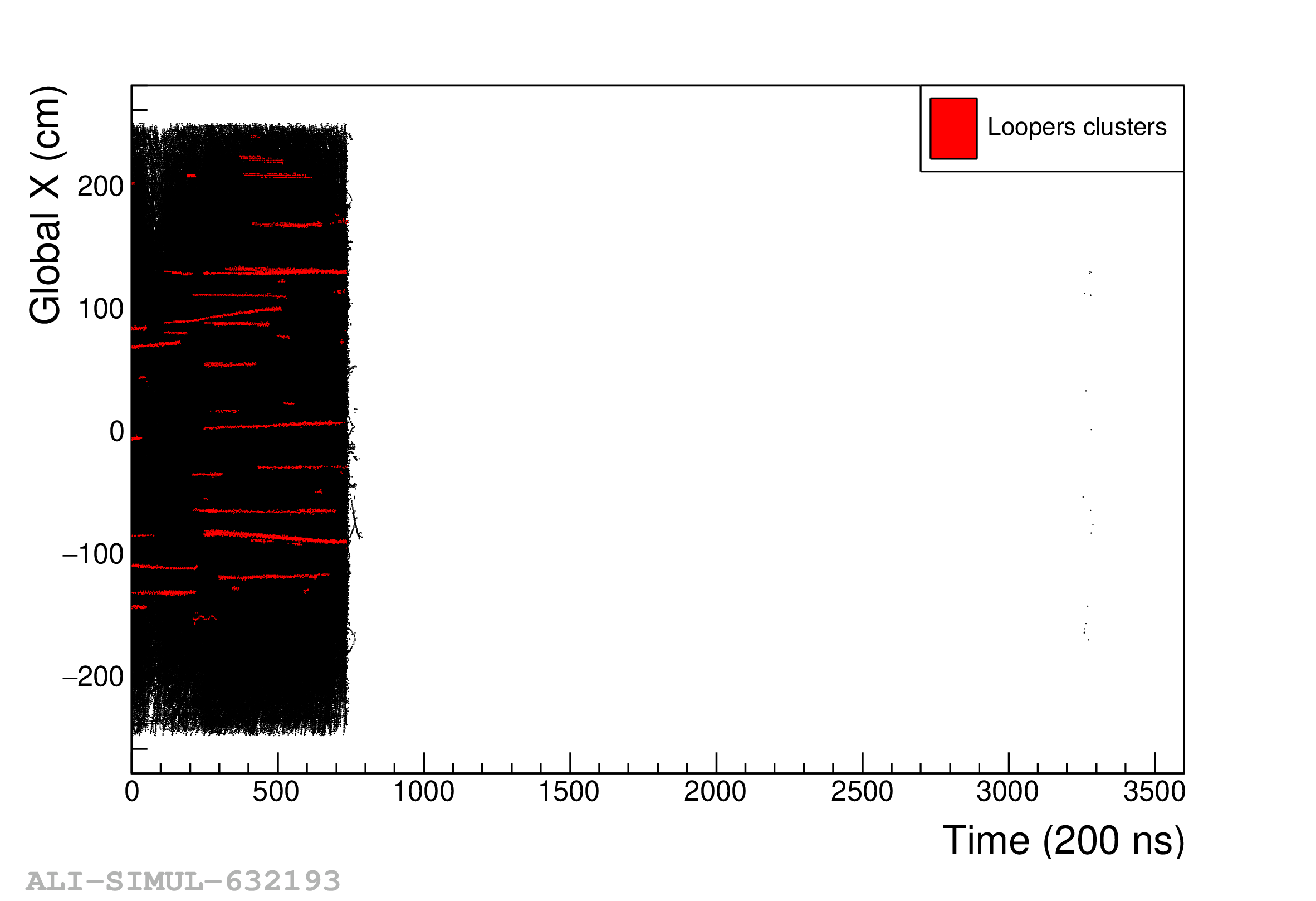}\hfill
\includegraphics[width=0.5\textwidth]{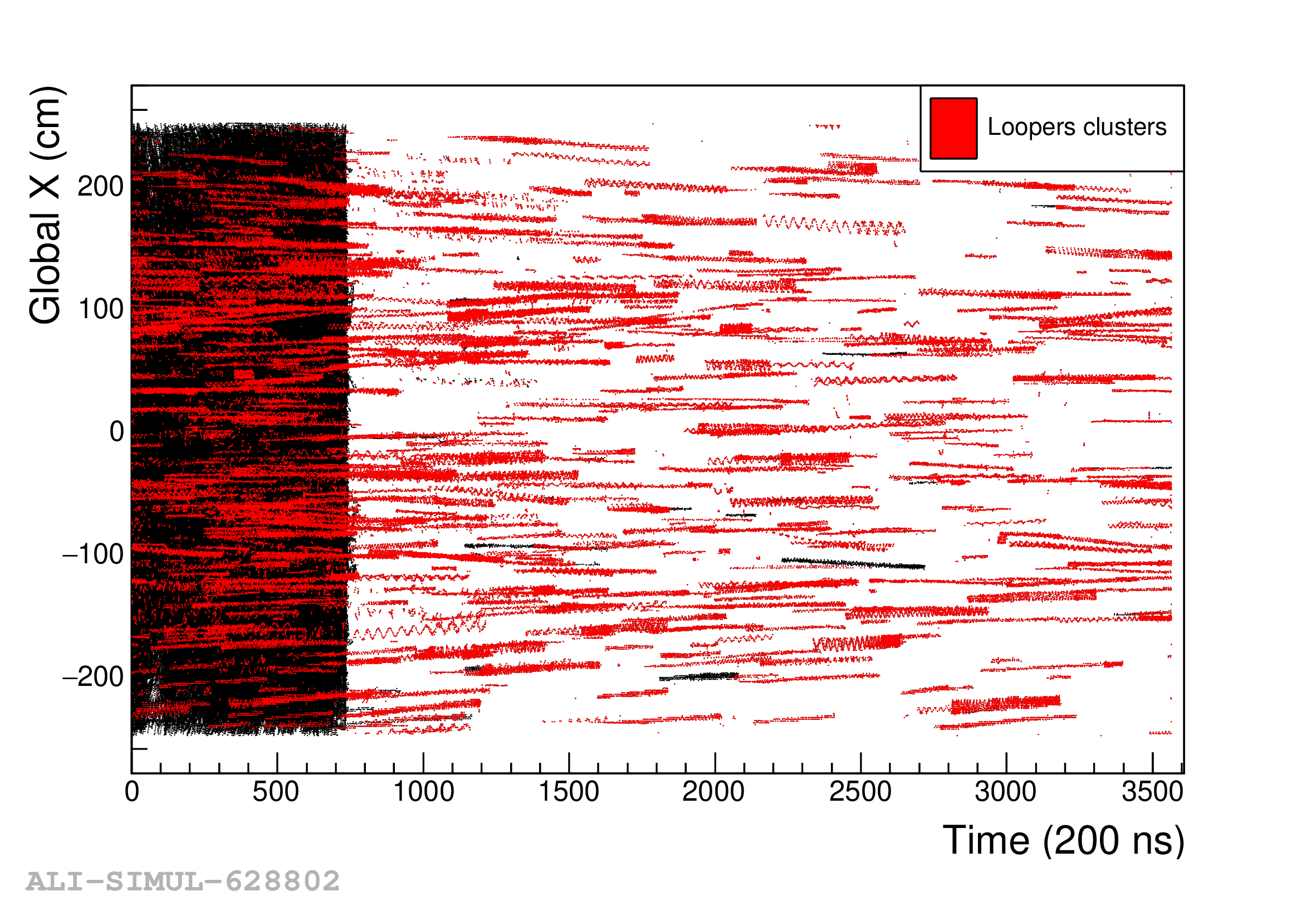}
\caption{Reconstructed TPC clusters (global $x$ position versus time in TPC units - 200 ns/bin)
in Pythia8+GEANT4 Pb--Pb simulations with the default slow-neutron
production cut (left) and with the relaxed cut used to build the
training sample (right); tagged looper clusters are shown in red.}
\label{fig-tagging}
\end{figure}

Loopers were then tagged offline in the resulting Monte Carlo record
by selecting particles satisfying the following criteria: (i) a
neutron mother with kinetic energy $K<5$~MeV, (ii) a $\gamma$
originating from the capture of that neutron, and (iii) either an
$e^+e^-$ pair from pair production or a single $e^-$ from Compton
scattering of that $\gamma$. Figure~\ref{fig-tagging} compares the
reconstructed TPC clusters obtained with the default production cuts
(slow $n$ OFF, left), where almost no loopers survive, to those
obtained with the relaxed cuts (slow $n$ ON, right), where a large
number of tagged looper clusters (highlighted in red) are visible.
The latter sample constitutes the training and validation dataset used
for the generative model described in the next section.

\section{Generative model}
\label{sec-ml}

\subsection{Choice of architecture}
\label{sec-ml-choice}
Three classes of generative models were evaluated for producing the kinematic and vertex features of looper particles.
Variational Auto-Encoders (VAEs)~\cite{Kingma:2019tpv}, widely used in
high-energy physics for anomaly detection and easy to implement, were found to poorly
capture the highly skewed distributions present in several looper
features: the high peaks in these distributions bias the learned
latent representation and shift the whole model, resulting in
unrealistic samples. Generative Adversarial Networks
(GANs)~\cite{Goodfellow:2014upx}, already used by other collaborations
for fast simulation, gave reasonable results but are prone to mode
collapse, where the generator concentrates on a limited subset of the
target distribution, and to unstable, non-converging training.

Finally, the Wasserstein GAN with gradient penalty
(WGAN-GP)~\cite{Arjovsky:2017fjr,Gulrajani2017wgangp} was found to
address both issues at the cost of a slightly more complex training procedure,
and was therefore adopted. In this method the discriminator is
replaced by a critic $D$ that assigns a continuous realism score to
generated samples rather than a bounded probability, mitigating the vanishing-gradient issue; the score is derived from the Wasserstein
distance between the generated and target distributions, providing
smoother and more informative gradients; and a gradient-penalty term
enforces the Lipschitz constraint required by the Wasserstein
formulation without the training instabilities introduced by weight
clipping, effectively making the critic a better teacher for the
generator. The critic and generator losses take the form
\begin{equation}
L = \mathbb{E}_{\tilde{\mathbf{x}}\sim\mathbb{P}_g}\!\left[D(\tilde{\mathbf{x}})\right]
  - \mathbb{E}_{\mathbf{x}\sim\mathbb{P}_r}\!\left[D(\mathbf{x})\right]
  + \lambda\,\mathbb{E}_{\hat{\mathbf{x}}\sim\mathbb{P}_{\hat{\mathbf{x}}}}\!
    \left[\left(\lVert \nabla_{\hat{\mathbf{x}}} D(\hat{\mathbf{x}}) \rVert_2 - 1\right)^2\right],
\label{eq-wgangp}
\end{equation}
where $\mathbb{E}$ defines the expected value, $\mathbb{P}_r$ and $\mathbb{P}_g$ are the real and generated
data distributions, $\mathbb{P}_{\hat{\mathbf{x}}}$ is the
distribution of points sampled along straight lines between real and
generated pairs, and $\lambda$ controls the strength of the gradient
penalty~\cite{Gulrajani2017wgangp}. Two independent WGAN-GP models are
trained for simplicity: one for $e^+e^-$ pair-production and one for
Compton-scattering electrons, since the two processes yield a different number of physical features.

\subsection{Network architecture}
\label{sec-ml-arch}
Both the generator and the critic are implemented as fully connected
feed-forward networks, summarised in
Table~\ref{tab-architecture}. The generator takes as input a
100-dimensional random noise vector and, through three fully
connected layers totalling 778 neurons, outputs the kinematic and
vertex features of the generated particles: seven for the
single-electron Compton model (three momentum components, three
vertex coordinates and the vertex time), and ten for the
pair-production model, which adds the three momentum components of
the second lepton. The critic
takes a feature vector (generated or from full simulation) as input
and, through four fully connected layers totalling 897 neurons,
outputs a single scalar realism score, with higher values indicating
greater similarity to the full-simulation training sample. During
training the generator and the critic are optimised alternately, each
improving in response to the other. The numbers of layers and neurons were chosen empirically,
as the smallest network that still describes the physical features of the particles well while keeping the inference time low. 

\begin{table}[h]
\centering
\begin{tabular}{lll}
\hline
 & Generator & Critic \\\hline
Input & noise (dim.\ 100) & features (7 or 10) \\
Layers & 3 fully connected & 4 fully connected \\
Neurons & 778 & 897 \\
Output & 7 or 10 features & realism score \\\hline
\end{tabular}
\caption{Summary of the WGAN-GP network architecture.}
\label{tab-architecture}
\end{table}

\section{Integration into the O2 simulation framework}
\label{sec-integration}
The generative models were developed and trained in Python; inference
was then implemented in C++ inside the ALICE O2 software
framework~\cite{Eulisse2019o2,Buncic:2011297} for use in production simulations.
Three new classes handle this integration. A scaler class applies a
combination of MinMax and robust scaling to the network input and
output features, with all scaling parameters parsed from a JSON
configuration file. An \texttt{ONNXGenerator} class loads network
checkpoints, translated from the PyTorch training format to the ONNX
format, and performs inference using ONNX
Runtime~\cite{OnnxRuntime}. Finally, a \texttt{LoopersGenerator} class
initialises the two classes above and translates the scaled network
output into \texttt{TParticle} objects. The network provides the
kinematics and vertex of a single looper but not their number, so the
absolute yield is set separately: the number of particles generated
per time interval is adjusted dynamically to give a flat looper
occupancy across the simulated timeframe, normalised to the rate
observed in full simulation. These particles are injected directly
into the GEANT4 transport stage, skipping entirely the transport of
the parent slow neutrons and their capture. This fast-simulation path
fits within the standard ALICE simulation workflow, from event
generation through detector transport, digitisation, and
reconstruction, to the analysis object data (AO2D) used by physics
analyses.

\section{Results}
\label{sec-results}
\begin{figure}[h]
\centering
\includegraphics[width=0.7\textwidth]{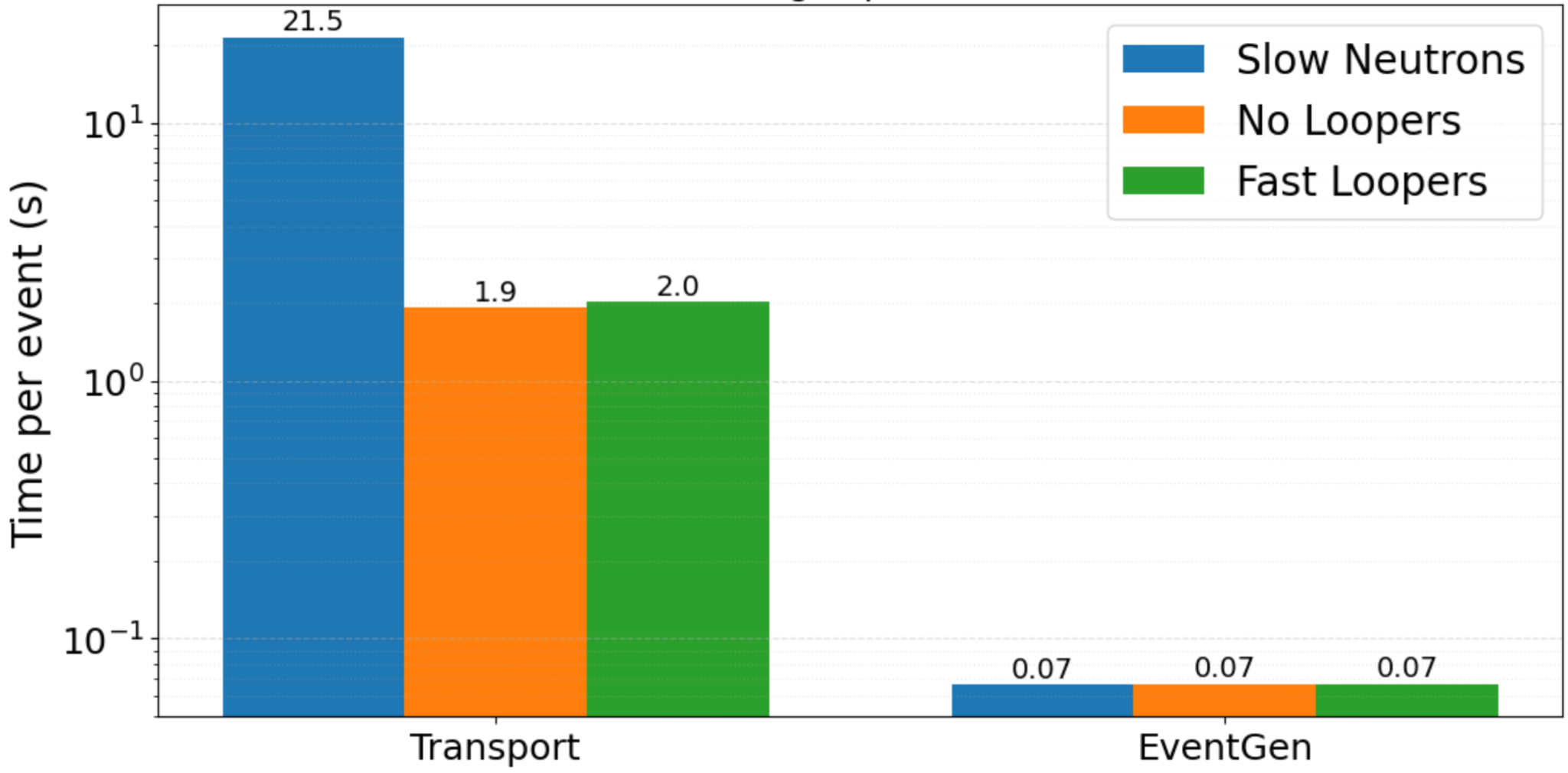}
\caption{Mean CPU time per event for the event-generation and
transport steps (single core test with Intel Xeon Gold 6132), comparing full slow-neutron transport, transport
with loopers entirely disabled, and transport with the WGAN-GP-based
fast looper generator. Results are based on pp collisions at $\sqrt{s}=13.6$~TeV with a 500kHz interaction rate.}  
\label{fig-speedup}
\end{figure}

\begin{figure}[!ht]
\centering
\includegraphics[width=0.7\textwidth]{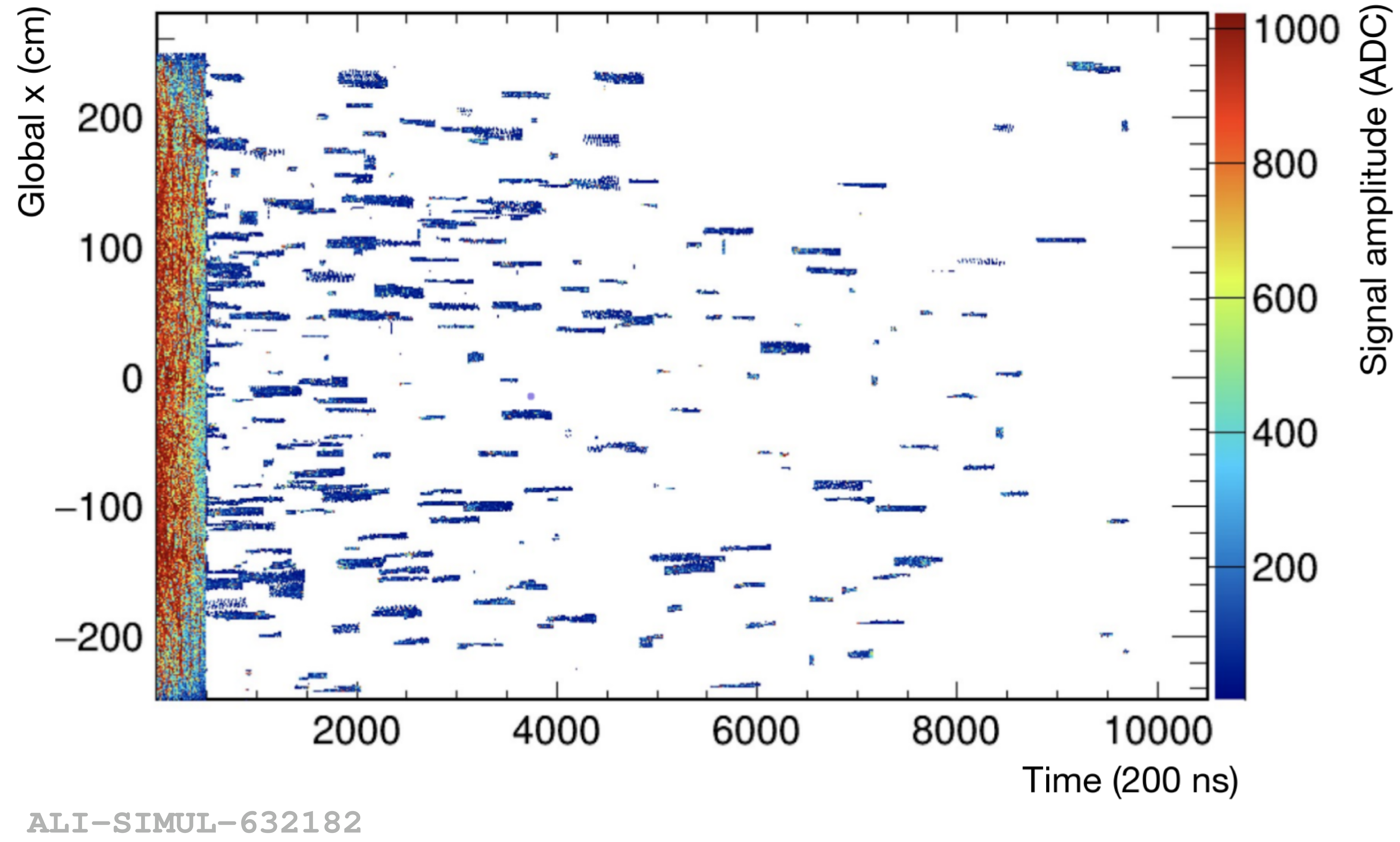}
\caption{Reconstructed TPC clusters (global $x$ position versus time in TPC units - 200 ns/bin)
from a Pb--Pb fast simulation using the WGAN-GP looper generator.}
\label{fig-fastsim-loopers}
\end{figure}

The fast looper generator has been extensively validated and is now
used in production for ALICE simulations in pp and Pb--Pb collisions. In
terms of computational performance, replacing the full slow-neutron
transport with the WGAN-based generator reduces the mean transport
time per event from 21.5~s to about 2.0~s, a speed-up of roughly a
factor 10, while leaving the event-generation time unaffected, as
shown in Fig.~\ref{fig-speedup}. Since transport with loopers disabled altogether
takes 1.9 s, the generator itself costs about 0.1 s per event and recovers almost all of the available saving.
\begin{figure}[!ht]
\centering
\includegraphics[width=\textwidth,clip]{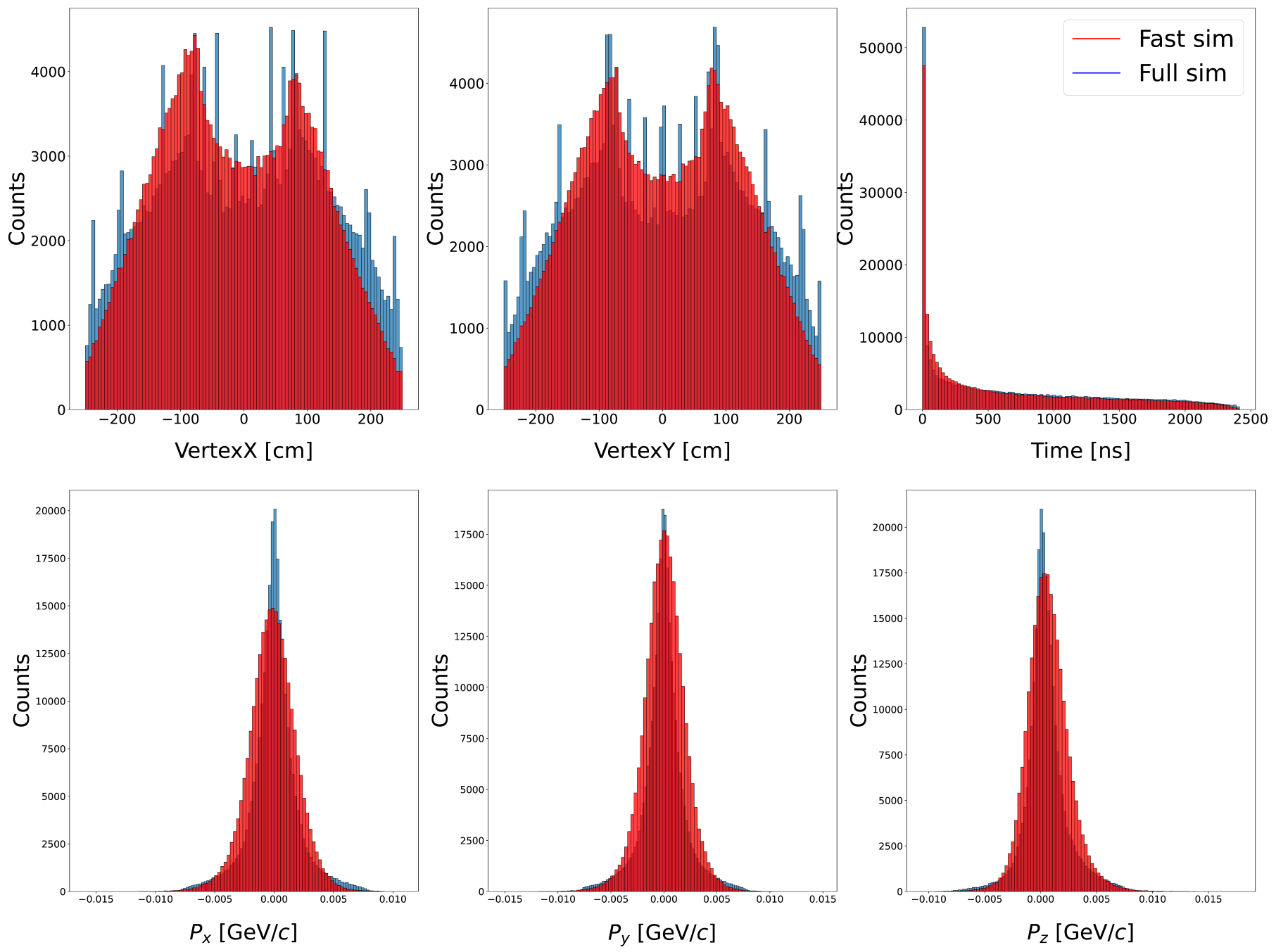}
\caption{Comparison between the WGAN-GP fast-simulation (red) and
full Pythia8+GEANT4 simulation (blue) distributions of vertex
position, time, and momentum components for Compton electrons.}
\label{fig-compton}
\end{figure}\newline
Beyond the computational speed-up, the physics features of the
generated particles were validated against the full simulation, as shown in Fig.~\ref{fig-compton}, which compares, for the Compton electrons, the
distributions of vertex position, timing, and momentum components
between the fast and full simulation. 
The generator reproduces the
overall shape of all distributions and the pair production model shows
a similar agreement. Some features, particularly the width of the
momentum tails, still leave significant room for improvement, and
alternative architectures are being explored (Sect.~\ref{sec-conclusions}).
Most importantly, the looper occupancy in the TPC is reproduced. 
The absolute yield is normalised as described in Sect.~\ref{sec-integration}, so what is tested here is the
topology and time structure of the loopers rather than their rate.
Figure~\ref{fig-fastsim-loopers} shows the reconstructed
clusters of a fast Pb--Pb simulation as a function of global $x$
position and time: the same long-lived, closely packed looper
structures seen in the full simulation of Fig.~\ref{fig-tagging} and in the experimental data of Fig.~\ref{fig-looper-sector} are reproduced.

\section{Conclusions and outlook}
\label{sec-conclusions}
A fast, machine-learning-based simulation of TPC loopers induced by
slow-neutron interactions has been developed for the ALICE experiment,
based on a Wasserstein GAN with gradient penalty. Trained on tagged
samples from full Pythia8+GEANT4 simulations, the model has been fully
integrated into the ALICE O2 simulation framework and is now running
in production for both pp and Pb--Pb collisions, providing results
consistent with full simulation, a measured order-of-magnitude
speed-up in the transport step, and correct reproduction of the TPC
occupancy due to loopers.

Alternative generative architectures are currently under study to
further improve both the generation speed and the accuracy with which
individual particle features are reproduced. In parallel, a fully
machine-learning-based approach to directly generate detector hits,
bypassing the GEANT4 transport of the generated leptons entirely, is
under test, and could yield further computational savings for this
and similar background processes.

%
\bibliography{references}

\end{document}